\documentclass[useAMS,usenatbib]{mnras} 

\usepackage{graphicx}
\usepackage{amssymb}
\usepackage{url}
\usepackage{multirow}
\usepackage{times} 
\usepackage{txfonts}
\usepackage{aas_macros} 

\usepackage[dvipsnames,usenames]{color}

\newcommand{\cd}{d$^{-1}$}
\newcommand{\kms}{km\,s$^{-1}$}

\title[Multi-instrument observations of HD\,42659]{HD\,42659: The only known roAp star in a spectroscopic binary observed with $B$ photometry, TESS and SALT\thanks{based on observations made with the Southern African Large Telescope (SALT)}}
\author[D. L. Holdsworth et al.]{Daniel L. Holdsworth$^{1}$\thanks{E-mail:dlholdsworth@uclan.ac.uk},
Hideyuki Saio$^{2}$ and Donald W. Kurtz$^{1}$\\
$^{1}$ Jeremiah Horrocks Institute, University of Central Lancashire, Preston PR1 2HE, UK\\
$^{2}$ Astronomical Institute, School of Science, Tohoku University, Sendai 980-8578, Japan
}

\date{Accepted 2019 August 23. Received 2019 August 22; in original form 2019 June 26}

\pubyear{2019}

\begin{document}
\label{firstpage}
\pagerange{\pageref{firstpage}--\pageref{lastpage}}
\maketitle

\begin{abstract}
We present a multi-instrument analysis of the rapidly oscillating Ap (roAp) star HD\,42659. We have obtained $B$ photometric data for this star and use these data, in conjunction with TESS observations, to analyse the high-frequency pulsation in detail. We find a triplet which is split by the rotation frequency of the star ($\nu_{\rm rot}=0.3756$\,\cd; $P_{\rm rot}=2.66$\,d) and present both distorted dipole and distorted quadrupole mode models. We show that the pulsation frequency, $150.9898$\,\cd\ ($P_{\rm puls}=9.54$\,min) is greater than the acoustic cutoff frequency. We utilise 27 high-resolution ($R\simeq65\,000$), high signal-to-noise ($\sim120$) spectra to provide new orbital parameters for this, the only known roAp star to be in a short period binary ($P_{\rm orb}=93.266$\,d). We find the system to be more eccentric than previously thought, with $e=0.317$, and suggest the companion is a mid-F to early-K star. We find no significant trend in the average pulsation mode amplitude with time, as measured by TESS, implying that the companion does not have an affect on the pulsation in this roAp star. We suggest further photometric observations of this star, and further studies to find more roAp stars in close binaries to characterise how binarity may affect the detection of roAp pulsations. 
\end{abstract}

\begin{keywords}
asteroseismology -- stars: chemically peculiar -- stars: oscillations -- techniques: photometric -- techniques: radial velocities -- stars: individual: HD\,42659
\end{keywords}

\section{Introduction}
\label{sec:intro}

The rapidly oscillating Ap (roAp) stars are a rare subclass of the Ap stars that show brightness and radial velocity variations with periods of $4.7-23.7$\,min. Discovered using targeted $B$ photometric observations of known Ap stars \citep{kurtz1978,kurtz1982}, only 76 roAp stars have been identified over a 40-yr period (see catalogues of \citealt{smalley2015,joshi2016} and recent works by \citealt{cunha2019,balona2019,hey2019}). With the recent results from the Transiting Exoplanet Survey Satellite \citep[TESS;][]{ricker2015}, it is becoming evident that the roAp stars are even rarer than previously thought, with an incidence rate amongst the Ap stars of about 4\,per\,cent \citep{cunha2019}.

The magnetic Ap stars are characterised by their chemical peculiarities, strong predominately dipolar magnetic fields and slow rotation. They constitute about 13\,per\,cent of the A stars as a whole \citep{wolff1968,wolff1983}. The magnetic fields in the Ap stars are inclined to the rotation axes, and can have magnetic field modulus strengths of up to about 35\,kG \citep{babcock1960,elkin2010}, which act to suppress convection, thus allowing gravitational settling of some elements and radiative levitation of others, most notably the rare earth elements. This phenomenon can result in overabundances, in the stellar surface layers, of rare earth elements, which often form spots at the magnetic poles of the Ap stars \citep{ryabchikova2007}. Those spots are stable over many decades to centuries, thus allowing for the accurate determination of the star's rotation period through modulation of its light curve, which can be of the order days to decades and more \citep{mathys2015}. It is also the magnetic field in the Ap stars that is thought to cause this slow rotation in many of them through magnetic braking \citep{stepien2000}.

The pulsations in the roAp stars, which are low-degree ($\ell$), high-overtone ($n$) pressure (p) modes, are thought to be driven by the $\kappa$-mechanism acting in the H\,{\sc{i}} ionization zone \citep{balmforth2001,cunha2002}. It was found that this mechanism is only viable when the magnetic field suppresses convection around the magnetic poles, explaining why it is the magnetic Ap stars that show these high-overtone pulsation modes that are not found in the non-magnetic A stars. This also provides an explanation as to why the pulsation axis in roAp stars is closely aligned to the magnetic axis. 

The $\kappa$-mechanism cannot explain the observed pulsation mode frequencies in some stars. In non-magnetic stars there is a theoretical limit to pulsation mode frequencies, the acoustic cutoff frequency, where the wavelength of the pulsation becomes short compared to the scale of the surface reflective boundary. However, there are many roAp stars that pulsate above this limit \citep[see figure 6 of][]{holdsworth2018b}. It was shown by \citet{cunha2013} that, in surface regions where convection is not suppressed by the magnetic field, turbulent pressure can be a viable excitation mechanism for the observed super-critical pulsations, i.e. pulsation mode frequencies that are observed above the theoretical acoustic cutoff frequency for a non-magnetic star. In addition, due to the strong magnetic fields in these stars, compressible Alfv\'en waves are present in the stellar atmosphere, with some, or even most, of the pulsation wave energy residing in the magnetic waves, thus providing the potential for  pulsation frequencies higher than the purely acoustic cutoff frequency \citep{sousa2008}.

As previously mentioned, the pulsation axis in roAp stars is misaligned with respect to the rotation axis, but closely aligned to the magnetic one. This can lead to a variation in the viewing aspect of a pulsation over the stellar rotation period. This phenomenon results in amplitude modulation of any pulsation mode with the rotation period, which manifests itself as a multiplet split exactly by the rotation frequency in a Fourier transform. This was first noted by \citet{kurtz1982} who developed the oblique pulsator model. Subsequently, that model has been revised and improved many times \citep[e.g.,][]{shibahashi1985a,shibahashi1985b,shibahashi1993,bigot2002,bigot2011}. This property of the roAp stars gives constraints on the mode geometry, since a multiplet of $2\ell+1$ components is expected for each mode, where the relative amplitudes of the multiplet components are dependent on the inclination of the rotation axis to the line of sight, $i_{\rm inc}$, and the obliquity of the magnetic axis with respect to the rotation axis, $\beta$.

The spectroscopic binary fraction of Ap stars is of the order 45\,\,per\,cent \citep{carrier2002,mathys2017}. However, the short period ($P<20$\,d) spectroscopic binary fraction amongst the Ap stars is very low, with only about 10 known \citep{landstreet2017}. Although some roAp stars have been observed to be in visual binary systems \citep{scholler2012}, there is only one known to be a spectroscopic binary: HD\,42659. 

\section{HD\,42659}
\label{sec:star}

The subject of this paper is HD\,42659 (UV\,Lep; TIC\,33601621); we present basic and derived properties for this star in Table\,\ref{tab:params}.

\begin{table}
\caption{Basic and derived stellar parameters of HD\,42659.}
    \centering
    \begin{tabular}{lrr}
    \hline
    Parameter & \multicolumn{1}{c}{Value} & Reference\\
        \hline
Spectral type & Ap\,SrCrEu & (1)\\
$m_V$ & $6.79\pm0.01$ & (2)\\
$M_V$ & $1.03\pm0.23$ &(3) \\
$\pi$ (mas) & $7.62\pm0.04$ & (4)\\
$d$ (pc) & $131.2\pm0.7$ & (4)\\
$\log T_{\rm eff}$ & $3.90\pm0.01$ & (3)\\
$\log (L/\rm{L_\odot})$ & $1.48\pm0.09$ & (3)\\
$v\sin i_{\rm inc}$ (\kms) & $19\pm1$ & (5) \\
$\langle B_z\rangle$ (kG)& $0.40\pm0.05$& (3,6)$^\dagger$\\
$M$ (M$_\odot$) & $2.10\pm0.10$ & (3)\\
$R$ (R$_\odot$) & $2.90\pm0.33$& From standard relation\\

         \hline
         \multicolumn{3}{l}{$^\dagger$Weighted mean. (1)\,\citet{houk1988};}\\
         \multicolumn{3}{l}{(2)\,\citet{hog2000}; (3)\,\citet{kochukhov2006};}\\
         \multicolumn{3}{l}{(4)\,\citet{GaiaDR2}; (5)\,\citet{elkin2008};}\\
         \multicolumn{3}{l}{(6)\,\citet{hubrig2006}.}
    \end{tabular}
    \label{tab:params}
\end{table}

HD\,42659 was discovered to be a pulsationally variable Ap star by \citet{martinez1993}. They observed the star through a $B$ filter for a single night with a photometer mounted on the 0.75-m telescope at the South African Astronomical Observatory (SAAO). Later, a more extensive photometric study was conducted by \citet{martinez1994}, where they observed the star for a further 11 nights on the 1.0-m SAAO telescope with the aim using a larger aperture to decrease scintillation noise in the amplitude spectrum. Their results showed variability on only a few nights, indicating strong amplitude modulation of the pulsation, with a period of a few days.

Using a data set combined of 5 nights, which had the highest duty cycle, they concluded that the pulsation had a frequency of 149.9472\,\cd\ (1.7355\,mHz), but stated that this value should be treated with caution due to the mode having low amplitude and the data a complex window pattern. Since those results, no further photometric observations were made.

Spectroscopic observations spanning 4-yr were obtained by \citet{hartmann2015} using the High-Accuracy Radial-velocity Planet Searcher (HARPS) spectrograph on the 3.6-m telescope at the European Southern Observatory (ESO). Their 48 spectra, obtained on 16 nights, revealed significant radial velocity (RV) variations which were successfully modelled as being caused by eccentric binary motion with $P_{\rm orb}=93$\,d . Although with a period greater than 20\,d, this system possesses the potential to study the interplay between the excitation, or suppression, of roAp pulsations by a binary companion.

\section{Photometry}
\label{sec:phot}

\subsection{The 2017 SAAO observations}
\label{sec:SAAO}

We conducted a three-week observing campaign from SAAO in 2017 December. Using the 1.0-m telescope and one of the Sutherland High Speed Optical Cameras \citep[SHOC;][]{coppejans2013}, we aimed to obtain the most complete light curve of HD\,42659 to date to provide robust values for the pulsation frequency and rotation frequency. The observations were conducted with a $B$ filter, where the signal-to-noise (S/N) ratio is greatest for the roAp stars \citep{medupe1998}. We provide a log of the observations in Table\,\ref{tab:log}.

\begin{table}
\caption{Log of 2017/2018 SAAO observations of HD\,42659. UTC date and BJD are given for the mid-point of the first exposure.  All observations were conducted with the SAAO 1.0-m telescope by DLH.}
    \centering
    \begin{tabular}{llcr}
    \hline
    Year & UTC date & BJD & \multicolumn{1}{c}{Duration}\\
        &           &  \multicolumn{1}{c}{(2450000.0+)}     & \multicolumn{1}{c}{(min)} \\
        \hline
        2017\\
        & Dec 12/13 & 8100.35792 & 260.2\\
        & Dec 13/14 & 8101.36511 & 48.3\\
        & Dec 14/15 & 8102.35343 & 381.9\\
        & Dec 15/16 & 8103.35379 & 341.9 \\
        & Dec 16/17 & 8104.38644 & 219.3 \\
        & Dec 19/20 & 8107.41714 & 273.5\\
        & Dec 20/21 & 8108.47363 & 30.9 \\
        & Dec 22/23 & 8110.32456 & 380.4\\
        & Dec 23/24 & 8111.31963 & 340.4\\
        & Dec 24/25 & 8112.31763 & 353.1\\
        & Dec 25/26 & 8113.31867 & 396.2\\
        & Dec 26/27 & 8114.30889 & 409.4\\
        & Dec 27/28 & 8115.31138 & 395.4\\
        & Dec 28/29 & 8116.30767 & 375.8\\
        & Dec 30/31 & 8118.29875 & 428.0\\
        & Dec 31/Jan 01 & 8119.30476 & 152.4 \\
        2018\\
        & Jan 01/02 & 8120.28947 & 457.5 \\
        & Jan 02/03 & 8121.29443& 448.7\\
  
         Total & & & 5693.4   \\
         \hline
    \end{tabular}
    \label{tab:log}
\end{table}

\begin{figure}
\includegraphics[width=\columnwidth]{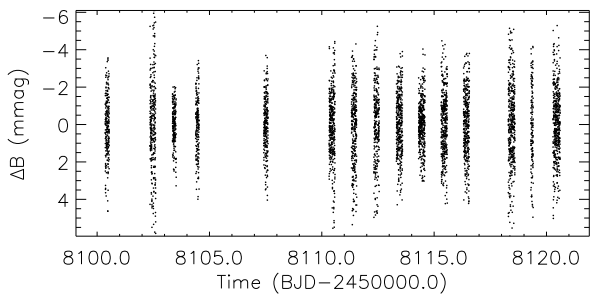}
\caption{Reduced 2017-2018 SAAO light curve for HD\,42659. Data affected by cloud have been removed, and pre-whitening has been performed to remove airmass and sky transparency variations.}
\label{fig:SAAO_lc}
\end{figure}

The data were reduced using the {\sc tea-phot} code of Bowman \& Holdsworth (submitted), which has previously been used for the reduction of SAAO/SHOC data \citep[e.g.,][]{holdsworth2018a}, and binned to 60-s integrations to give each data point the same weighting. Due to the brightness of the star ($V=6.8$) and the relatively small field of view of the SHOC instrument, there were no suitable comparison stars to perform differential photometry. We therefore removed data that was affected by cloud, then filtered low-frequency variations by pre-whitening the data. This process removed both airmass and sky transparency variations, but also any information on the rotation period of the star. The final light curve is shown in Fig.\,\ref{fig:SAAO_lc}.

\begin{figure}
\includegraphics[width=\columnwidth]{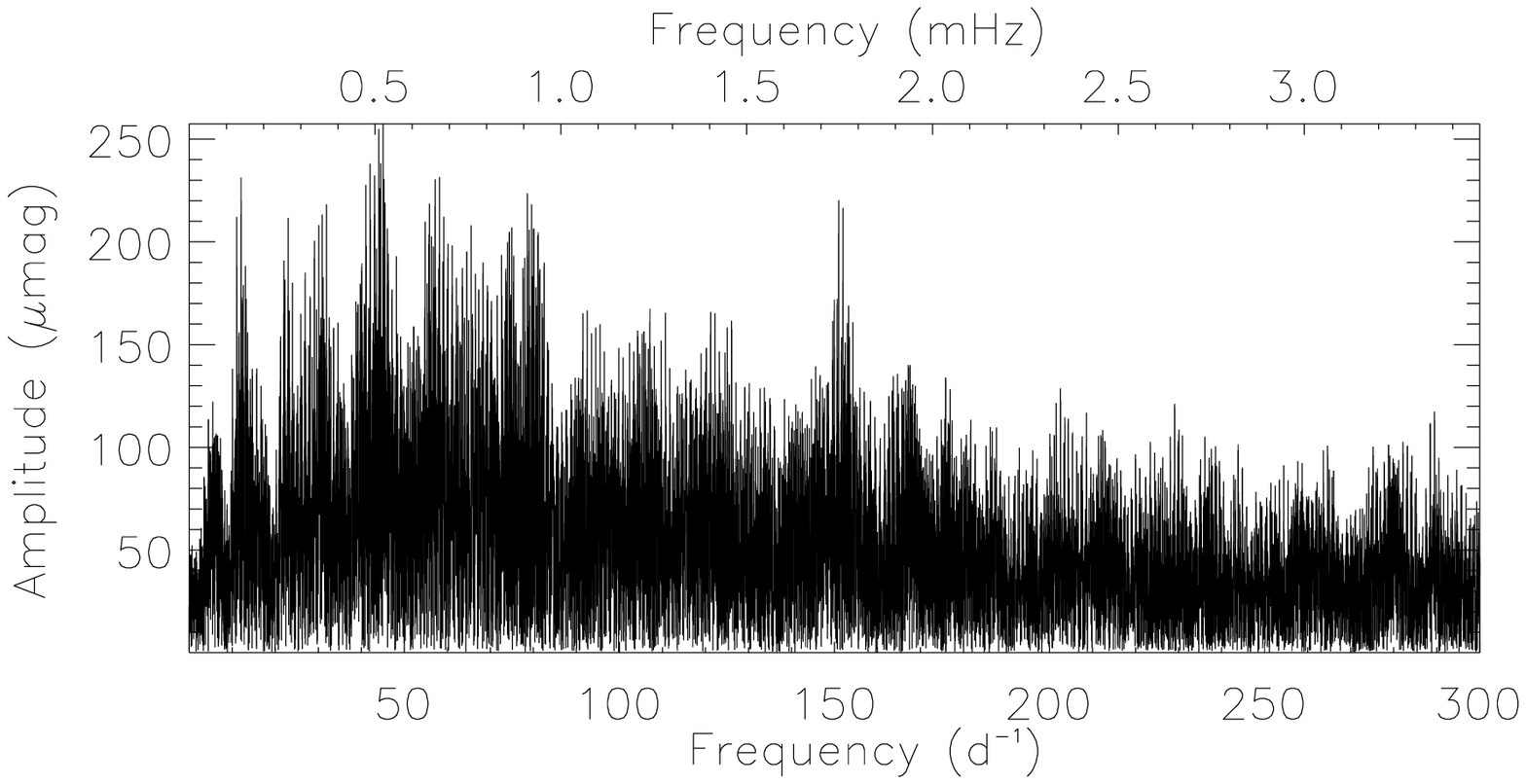}
\includegraphics[width=\columnwidth]{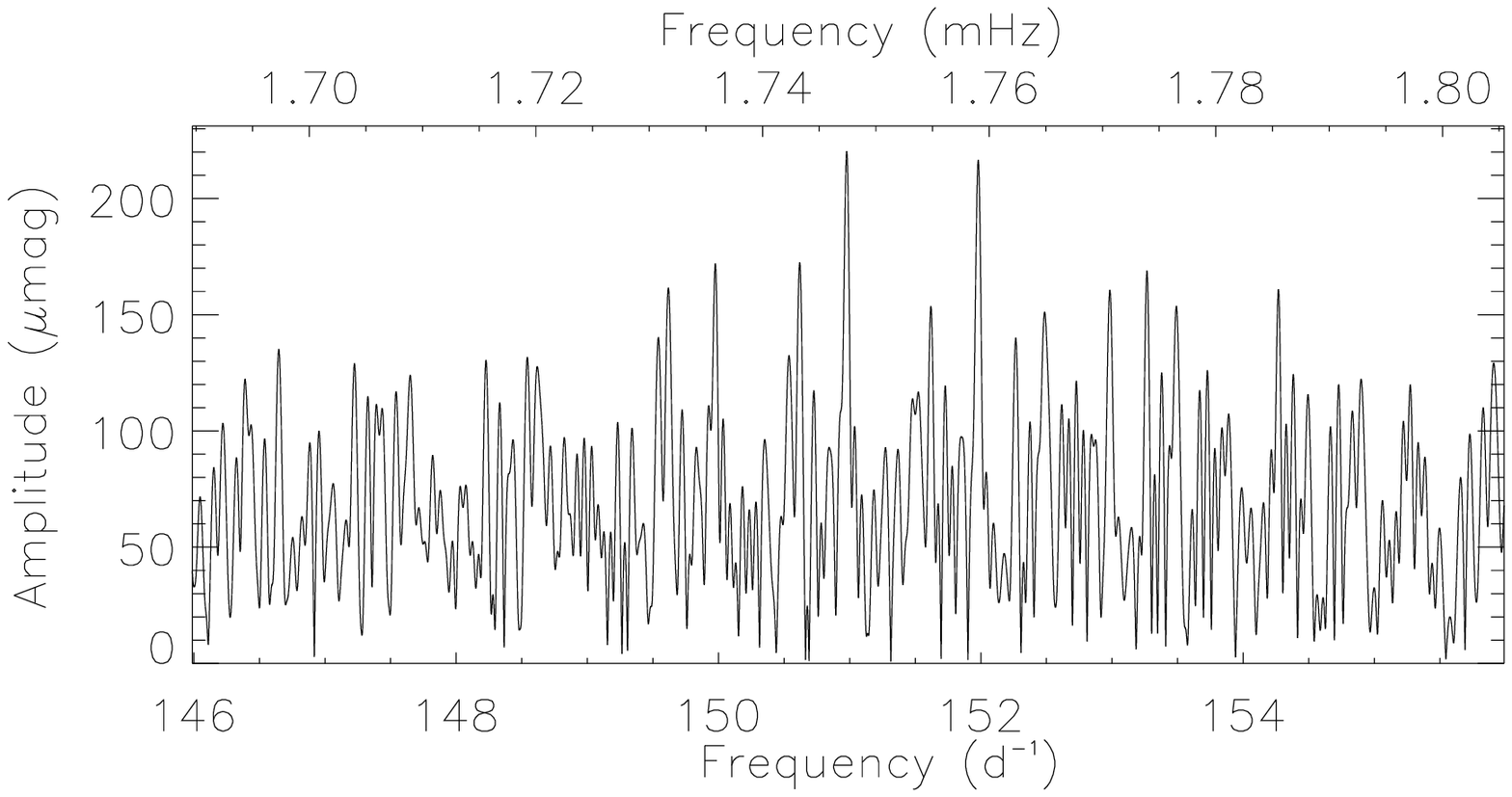}
\caption{Top: amplitude spectrum of the 2017-2018 SAAO observations of HD\,42659. There is a large amount of noise remaining in the light curve, as demonstrated by the excess power up to about 100\,\cd. Apparently evident is the pulsation frequency at 150\,\cd. Bottom: zoom of the suspected detection of the pulsation frequency. The highest amplitude peak falls at 150.98\,\cd\ but the second peak at 151.98\,\cd\ is of almost the same amplitude. We cannot determine which, if either, of these peaks is the true pulsation mode with these data.}
\label{fig:SAAO_ft}
\end{figure}

In search of the pulsation in HD\,42659, we calculate an amplitude spectrum to 300\,\cd\ and show this in the top panel of Fig.\,\ref{fig:SAAO_ft}. Despite efforts to remove the low-frequency artefacts in the data, the amplitude spectrum shows there is still a significant amount of red noise present. However, at higher frequency where the noise decreases, there is a clear feature which is significant above the noise. A detailed look (bottom panel of Fig.\,\ref{fig:SAAO_ft}) shows that there are two peaks in the amplitude spectrum ($\nu_a= 150.971\pm0.007$\,\cd, $\nu_b= 151.986\pm0.008$\,\cd), separated by 1\,\cd, which are in the same frequency range as the peak identified in the discovery data \citep{martinez1993}. With the 2017-2018 SAAO data alone, it is not possible to distinguish which of the peaks represents the pulsation mode. Furthermore, given the low S/N of this data set, no extra information can be gained from these data. What we have been able to show here is that the frequency previously published for this star, $149.94$\,\cd\ (1.7355\,mHz), is probably incorrect as a result of daily aliasing, however the authors did exercise caution in their conclusion \citep{martinez1994}.

\subsection{TESS observations}
\label{sec:TESS}

By far the best photometric data set for HD\,42659 was obtained by the TESS mission during sector 6. During its 2-yr mission, TESS will survey almost the entire sky with a cadence of 30\,min \citep{ricker2015}. The spacecraft consists of 4 cameras in a configuration which images a strip of sky $24^\circ\times96^\circ$ in size. One camera is centred on the ecliptic pole, while the others extend nearly to the ecliptic plane. The observing strategy consists of 13 sectors per hemisphere which are rotated through every $\sim27$\,d. As such, objects near the ecliptic pole will be observed for about 350\,d with objects near the ecliptic plane having just 27-d of data acquired \citep{ricker2015}.

In addition to the 30-min data, 20\,000 stars per sector were chosen to be observed in the high-cadence mode, with observations every 2~min. HD\,42659 was included in the 2-min cadence sample along with nearly 1400 other chemically peculiar stars in the context of the roAp programme of the TESS Asteroseismic Science Consortium (TASC). It is those high-cadence data that we use here; they are publicly available and obtainable from the Mikulski Archive for Space Telescopes (MAST\footnote{{\url{https://archive.stsci.edu}}}) website.

The data for HD\,42659 cover a time span of 21.77\,d, with a $1.09\,$d gap for data download. For the analysis, we used the PDC\_SAP fluxes, which were produced by the Science Processing Operations Centre \citep[SPOC;][]{jenkins2016}. The data were converted to magnitudes, and time stamps adjusted for the zero-point correction. The data were of good quality, thus we did not remove any outlying points. We did, however, remove low-frequency instrumental artefacts that remained after the pipeline reductions (a low-frequency trend and some residual background flux which were removed by fitting a series of sine curves). An initial analysis of these TESS data was presented by \citet{balona2019}, but we provide a more detailed analysis here.

\begin{figure}
\includegraphics[width=\columnwidth]{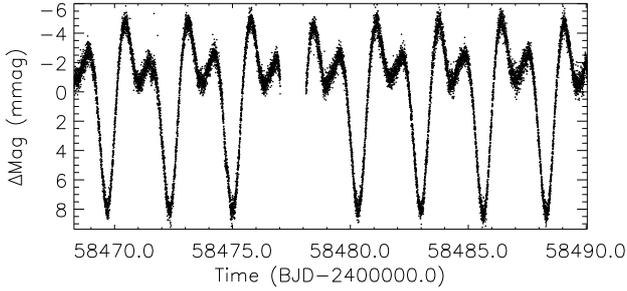}
\caption{Full TESS light curve of HD\,42659. There is a clear rotation signature due to chemical spots. We have removed some instrumental artefacts from the data.}
\label{fig:lc}
\end{figure}

\begin{figure}
\includegraphics[width=\columnwidth]{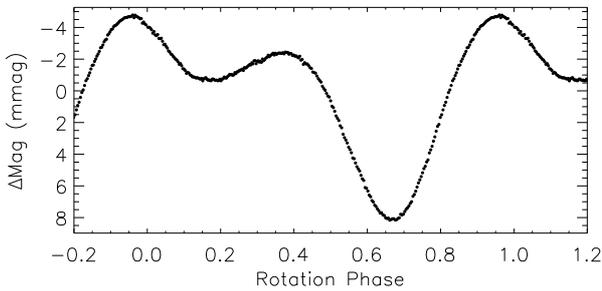}
\caption{Phase folded light curve showing the stability of the spots over the observing period, as expected. The data have been binned 50:1, and the zero point was chosen to be the time of pulsation maximum, BJD\,2458481.21190.}
\label{fig:rot}
\end{figure}

At first visual inspection, there is a clear modulation in the light curve of HD\,42659 (Fig.\,\ref{fig:lc}). This is expected for the Ap stars, which host spots of increased chemical abundances. To determine the rotation period of the star, we calculated the amplitude spectrum using a discrete Fourier transform \citep{kurtz1985} and fitted a harmonic series by non-linear least-squares to the data. A total of six frequencies were used such that the the amplitude of the highest frequency is at approximately the noise level in the data. We derived a rotation frequency of $0.375566\pm0.000035$\,\cd, which corresponds to a period of $2.66265\pm0.00025$\,d. We show a phase folded light curve plot in Fig.\,\ref{fig:rot}.

\begin{figure}
\includegraphics[width=\columnwidth]{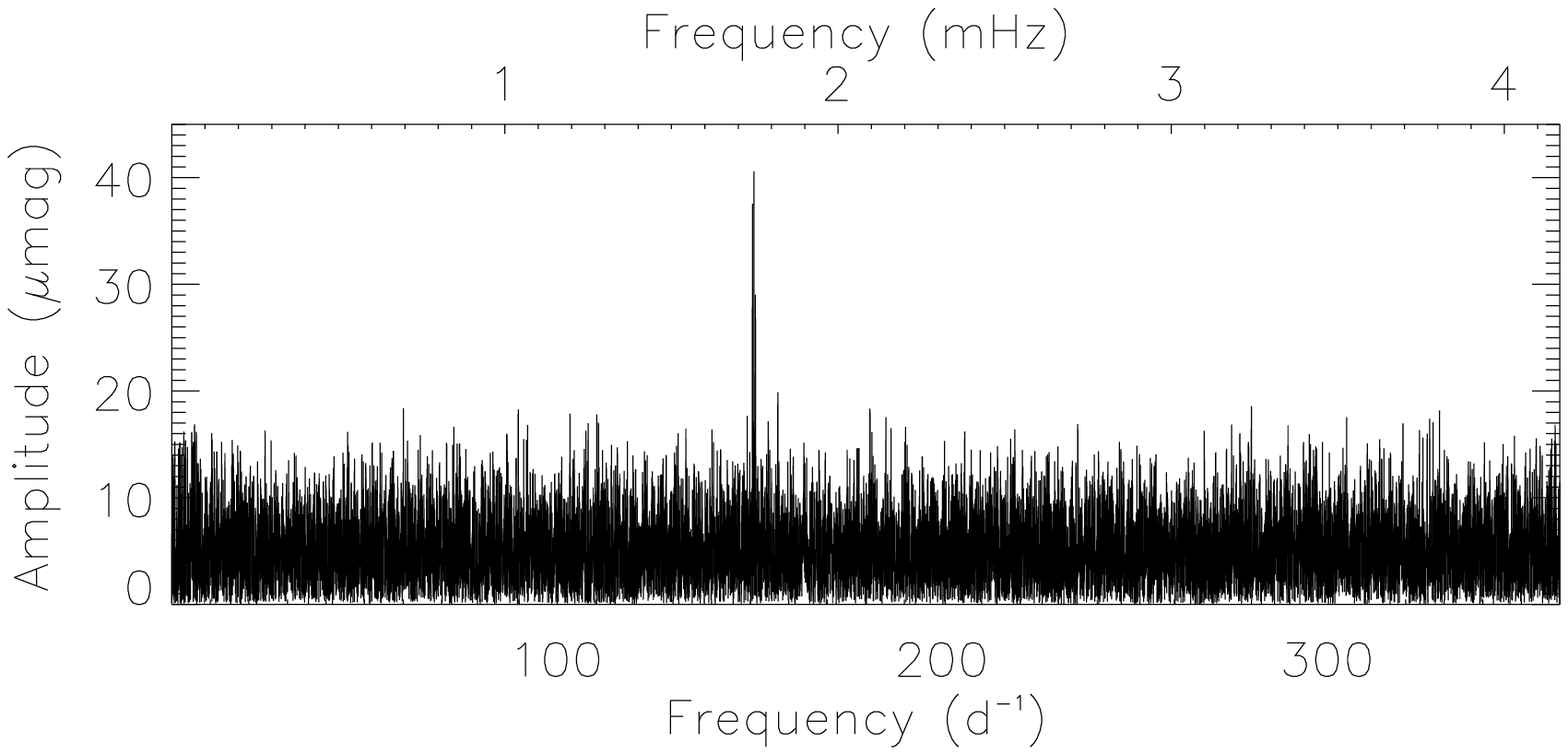}
\includegraphics[width=\columnwidth]{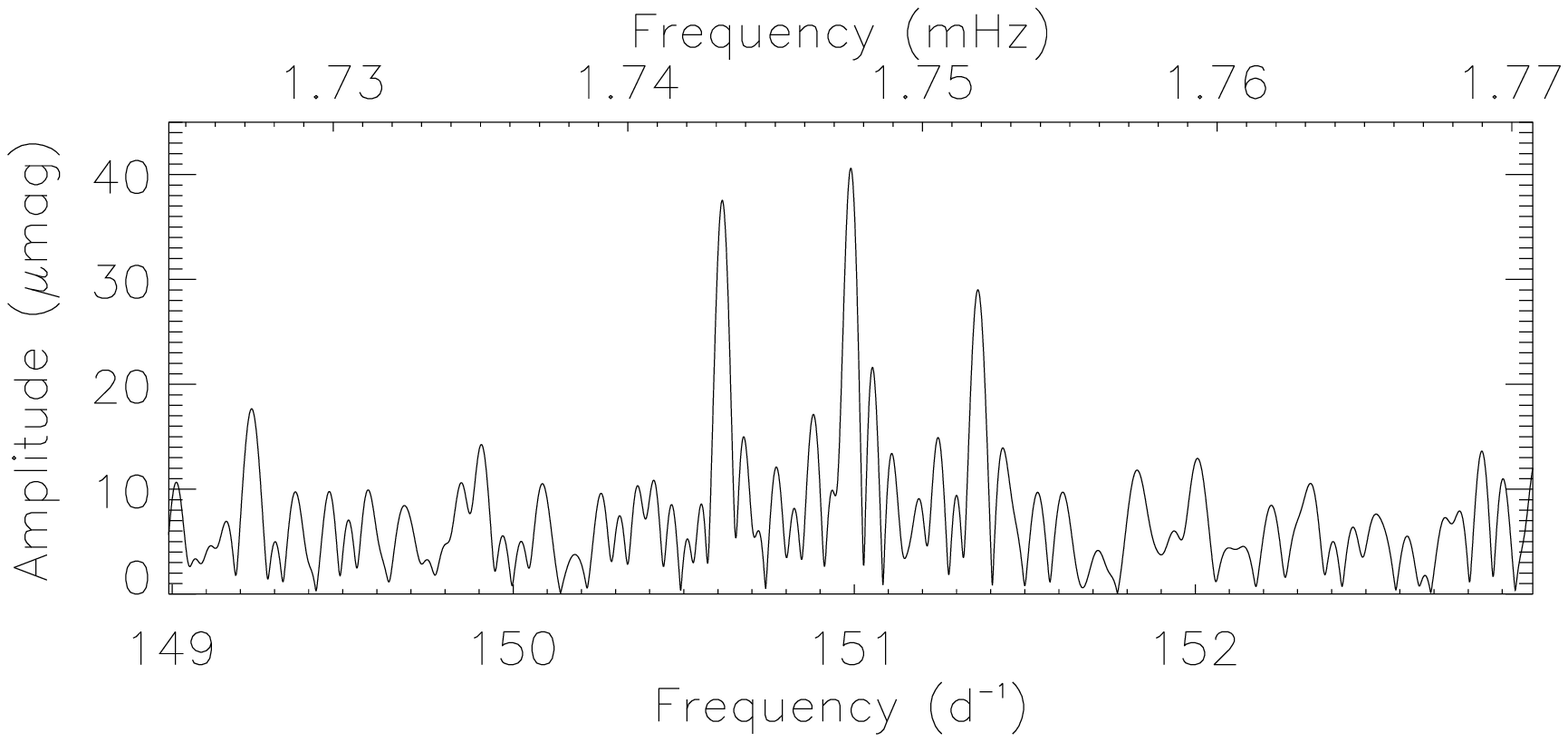}
\caption{Top panel: the full amplitude spectrum of HD\,42569 to the Nyquist frequency of the TESS data. Bottom: zoomed view of the pulsation mode frequencies. Clearly the mode is split into a triplet, with the components split by the rotation frequency.}
\label{fig:ft}
\end{figure}

To analyse the pulsation in the TESS data, we pre-whitened the data in the range $0-20$\,\cd\ ($0-0.23$\,mHz) to an amplitude level that is approximately that of the noise at high-frequency (i.e., the average peak height in the range $200-300$\,\cd as shown in Fig.\,\ref{fig:ft}) . Before performing this step, we checked that there is no other signal in this range that could be attributed to an astrophysical signal. After removing the rotation harmonic series, there are only low-amplitude instrumental artefacts remaining. 

The pulsation is clearly evident in the top panel of Fig.\,\ref{fig:ft}, with a more detailed view in the bottom panel. The multiplet is indicative of a dipole mode, as under the oblique pulsator model, one expects a multiplet of $2\ell+1$ peaks for a given mode. These high precision TESS data allow us to confirm that the pulsation frequency in this star is $150.9898\pm0.0029$\,\cd ($1.747567\pm0.000034$\,mHz), which is consistent with the highest peak found in the 2017 SAAO data set.

To analyse the multiplet, we iteratively fitted the components non-linear least-squares. We show the results of this fitting procedure in Table\,\ref{tab:nnls}. After removing the multiplet from the data, there are no further significant peaks remaining. We used this non-linear fit to show that the splitting is equal to the rotation frequency by taking the ratio of the sidelobe splitting to the rotation frequency presented above. For the $\nu-\nu_{\rm rot}$ sidelobe we found the ratio to be $1.004\pm0.011$ and for the $\nu+\nu_{\rm rot}$ the ratio was $1.008\pm0.013$, both, in agreement with the rotation frequency derived above.

\begin{table}
\caption{The results of a non-linear least-squares fit to the pulsation multiplet. The zero point for the phases is BJD\,2458479.16258.}
    \centering
    \begin{tabular}{lccr}
    \hline
    ID & Frequency & Amplitude & \multicolumn{1}{c}{Phase} \\
       & (\cd)     &($\umu$mag) & \multicolumn{1}{c}{(rad)}\\
       \hline
       $\nu-\nu_{\rm rot}$ & $150.6128\pm0.0030$ & $ 	36.8\pm4.5$ & $0.97\pm0.12$\\
       $\nu$               & $150.9898\pm0.0029$ & $	38.7\pm4.5$ & $2.44\pm0.12$\\
       $\nu+\nu_{\rm rot}$ & $151.3625\pm0.0039$ & $	28.4\pm4.5$ & $-2.42\pm0.16$\\
\hline
    \end{tabular}
    \label{tab:nnls}
\end{table}

Returning to the oblique pulsator model, for a pure, non-distorted mode, one expects, for the right choice of time zero point, the phases of all components of the frequency multiplet to be equal; this is a signature of pure amplitude modulation. To test this, we assumed that the sidelobes to the pulsation frequency are split from that frequency by exactly the rotation frequency of the star. Therefore, using the rotation frequency derived above, we force-fitted the sidelobes, then fitted the light curve by linear least-squares, choosing a zero point in time such that the phases of the sidelobes are equal. The results of this test are shown in Table\,\ref{tab:lls}. 

\begin{table}
\caption{The results of a linear least-squares fit to the pulsation multiplet where we forced the sidelobes to be split from the pulsation frequency by exactly the rotation frequency. The zero point for the phases is BJD\,2458481.21190.}
    \centering
    \begin{tabular}{lccr}
    \hline
    ID & Frequency & Amplitude & \multicolumn{1}{c}{Phase} \\
       & (\cd)     &($\umu$mag) & \multicolumn{1}{c}{(Rad)}\\
       \hline
$\nu-\nu_{\rm rot}$ & $150.6142$ & $36.8\pm4.5$ & $-1.18\pm0.12$ \\
$\nu$               & $150.9898$ & $38.7\pm4.5$ & $-1.17\pm0.12$ \\
$\nu+\nu_{\rm rot}$ & $151.3654$ & $28.2\pm4.5$ & $-1.18\pm0.16$ \\
\hline
    \end{tabular}
    \label{tab:lls}
\end{table}

It is evident that the phases are in excellent agreement, showing that the multiplet is a result of pure amplitude modulation, a signature of oblique pulsation. The amplitudes of the sidelobes, although unequal as a result of the Coriolis force \citep[e.g.,][]{shibahashi1993,bigot2002}, are in agreement within the errors. It seems that HD\,42659 is a good example of a `well behaved' roAp star. This makes the application of the full oblique pulsator model to this star valid.

\begin{figure}
\includegraphics[width=\columnwidth]{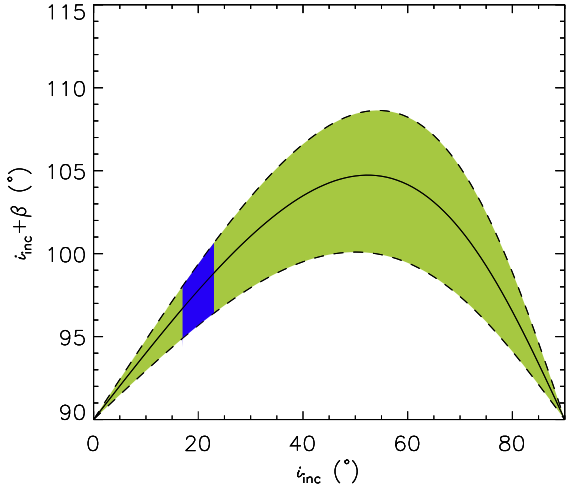}
\caption{Visualisation of the values of $i_{\rm inc}$ and $\beta$ that satisfy equation\,(\ref{eq:OPM_trip}). The green shaded region bound by the dashed lines represents the $1\sigma$ error range. As can be seen $i_{\rm inc}+\beta$ is always greater than $90^\circ$, suggesting we see both pulsation poles. The blue shaded region represents the values of $i_{\rm inc}$, which are estimated from spectroscopy. See the discussion for details.}
\label{fig:i_beta}
\end{figure}

In the absence of limb darkening and spots, \citet{kurtz1990} provided a relationship between the amplitudes of a dipole triplet, and the geometry of the star:

\begin{equation}
\label{eq:OPM_trip}
    \tan i_{\rm inc}\tan\beta = {\frac{A_{+1}^{(1)}+A_{-1}^{(1)}}{A_0^{(1)}}}, 
\end{equation}
where $A^{(1)}_{\pm1}$ are the sidelobe amplitudes, $A_0$ is the amplitude of the pulsation mode, $i_{\rm inc}$ and $\beta$ are the angles of inclination and magnetic obliquity, respectively. Using the results from Table\,\ref{tab:nnls} we find that $\tan i\tan\beta =1.68\pm0.26$. Although, without direct information on either $i_{\rm inc}$ or $\beta$, we are only able to provide constrains on the two angles, which are shown in Fig.\,\ref{fig:i_beta}.

We show the sum of $i_{\rm inc}$ and $\beta$ values as these enable visualisation of the star. For a pure dipole mode, the equator is a node and both poles are anti-nodes. Given that we see both magnetic poles, as suggested by the double-wave nature of the rotational variation of the light curve, it is likely we see both pulsation poles. For this to be the case, $i_{\rm inc}+\beta$ must be $>90^\circ$, which is satisfied by the results of equation\,(\ref{eq:OPM_trip}) and Fig.\,\ref{fig:i_beta}. In this case, the equatorial node crosses the line of sight and the amplitude should go to zero and a phase change of $\pi$-rad is expected.

\begin{figure}
\includegraphics[width=\columnwidth]{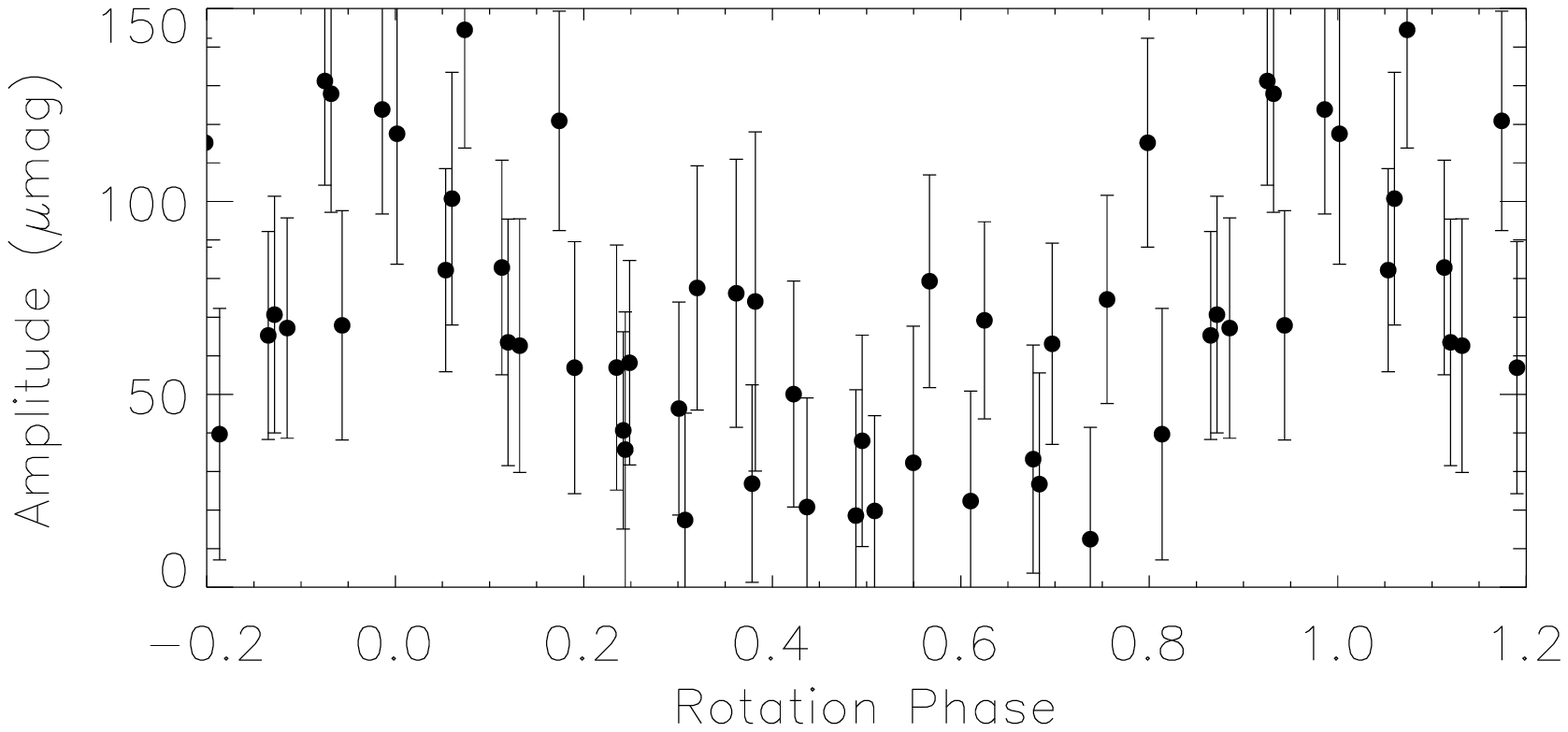}
\includegraphics[width=\columnwidth]{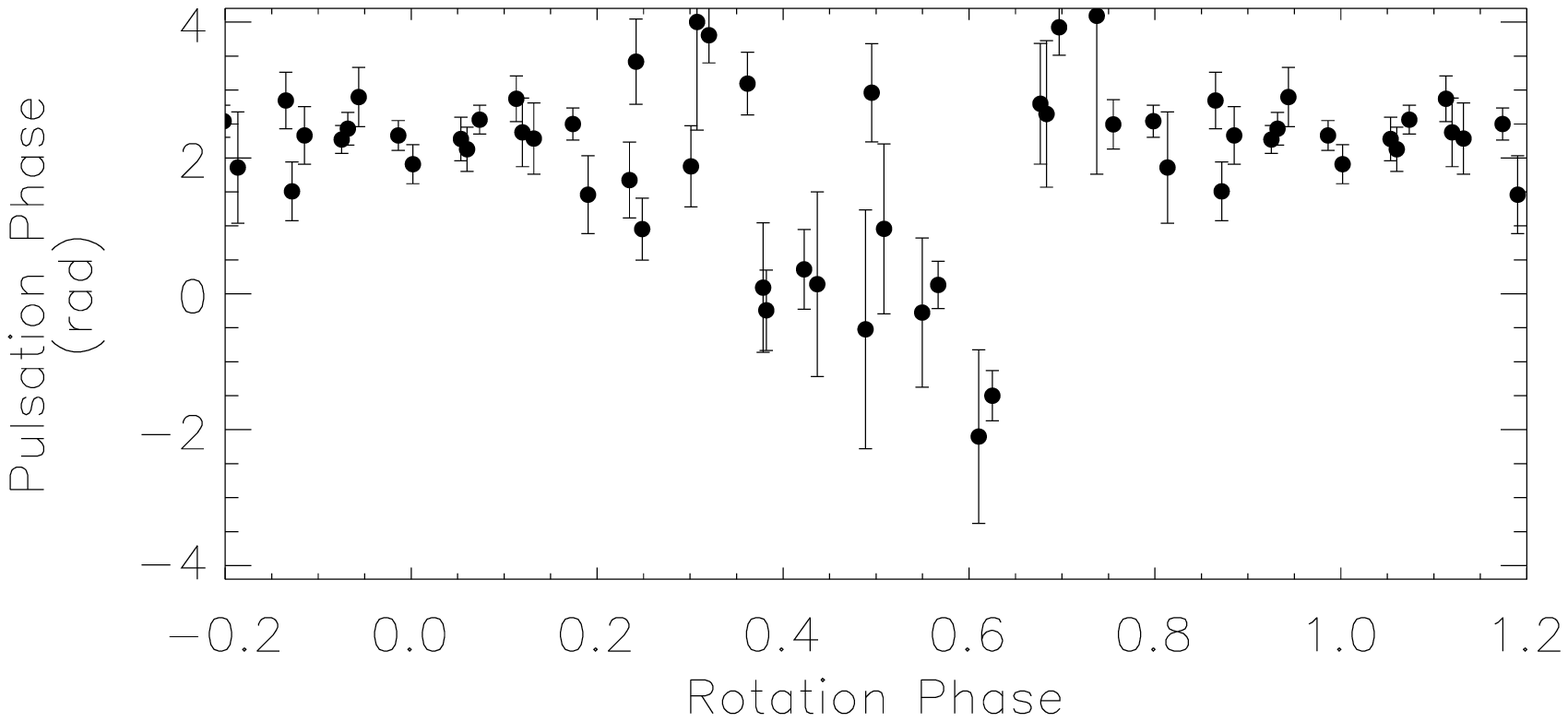}
\caption{Top panel: the variation in the pulsation amplitude over the rotation period of the star. Bottom: corresponding phase change of the pulsation mode. }
\label{fig:phamp}
\end{figure}

To test this, we split the light curve into discrete segments of 0.5\,d and calculated the amplitude and phase at fixed frequency, with the results shown in Fig.\,\ref{fig:phamp}. Although the scatter is large, due to the low mode amplitude, there is an obvious change in the pulsation amplitude and phase over the rotation period. This is consistent with the oblique pulsator model.

In conducting this test, we noticed an interesting phenomenon. Although the choice of segment length is arbitrary, as long as there is sufficient frequency resolution, we often choose integer multiples of the pulsation period. We did this and found, when phasing the data on the rotation period, our amplitude and phase measurements all fell at discrete rotation phases. Further investigation showed that the rotation period is $401.94\pm0.06$ times the pulsation period, i.e. an integer multiple within the errors. Whether this is coincidence or related is yet to be determined.

\begin{figure}
\includegraphics[width=\columnwidth]{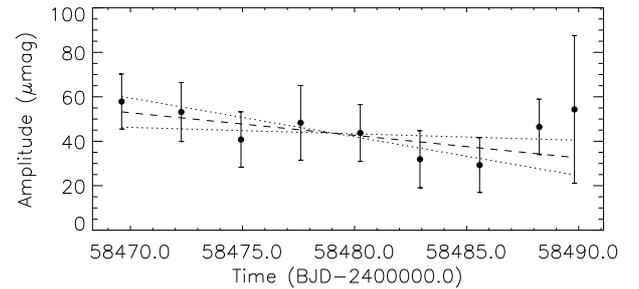}
\caption{The variation of pulsation amplitude over the length of the TESS observations. The large error on the final data point is a result of few data points being used for the amplitude calculation. The dashed line represents a linear fit to the data, with the dotted lines showing the $1\,\sigma$ confidence limits of the fit.}
\label{fig:rot_amp}
\end{figure}

Furthermore, the high scatter in the amplitude plot in Fig.\,\ref{fig:phamp} aroused suspicion, as one would expect the pulsation amplitude to be similar at the same phase for different rotation cycles (cf. figure 15 in \citealt{holdsworth2016}). To investigate this further, we calculated the pulsation amplitude at a fixed frequency in segments of the same length of the rotation period. In this way, we sample the same average pulsation amplitude over one complete rotation to the next. The results, shown in Fig.\,\ref{fig:rot_amp}, suggest that the average pulsation amplitude may not be stable over the 21-d TESS observations. We therefore fitted a linear regression to the data and found a gradient of $-1.0\pm0.7\,\muup$mag\,\cd, with a p-value of $0.32$ and a $\chi^2$ of $2.26$. This fit showed that there is no statistically significant change to the pulsation amplitude over the TESS observations period. We attribute the scatter in Fig.\,\ref{fig:phamp} to the low S/N of the data.

\begin{figure*}
\includegraphics[width=0.97\columnwidth]{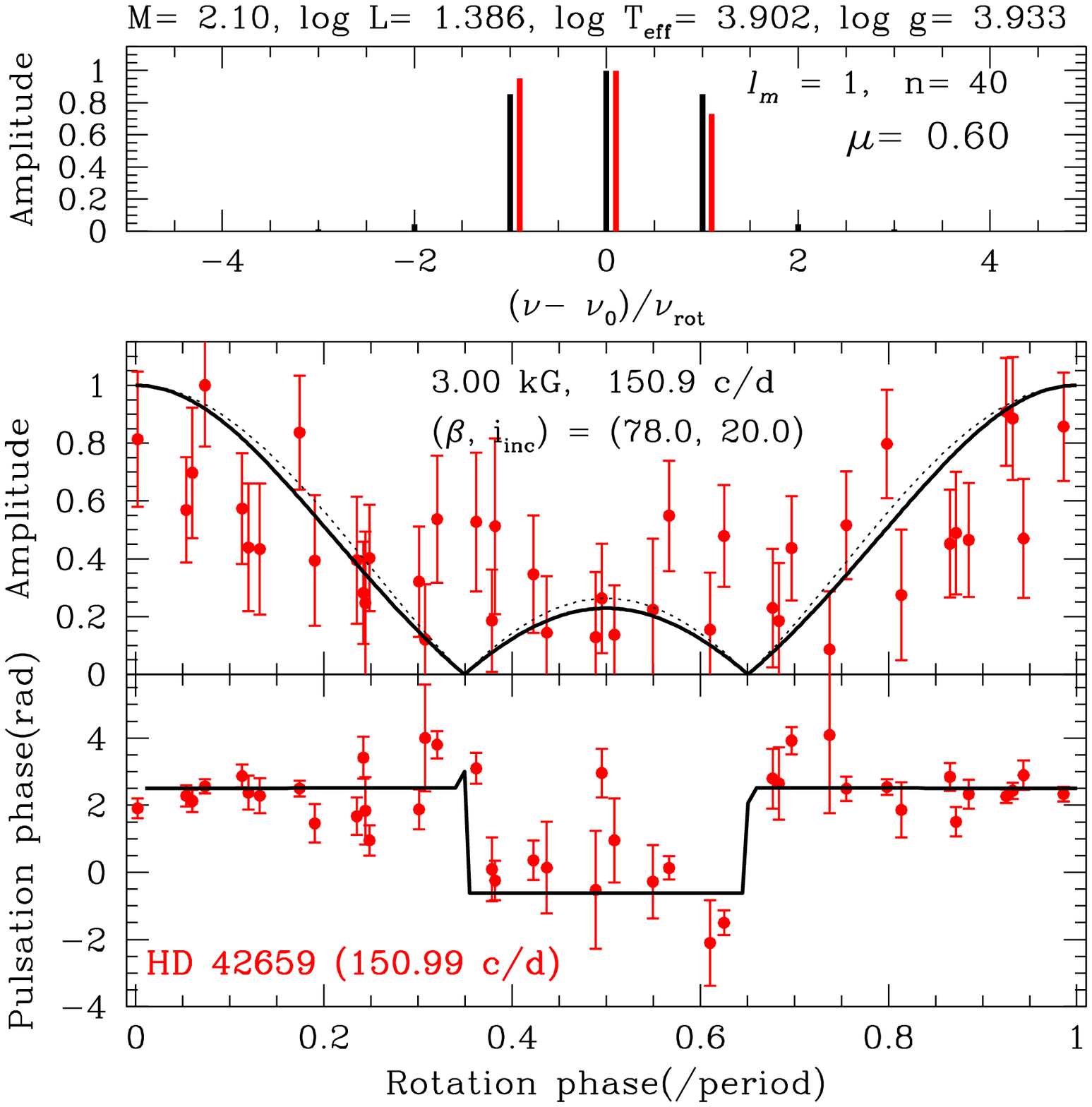}\hfill
\includegraphics[width=0.97\columnwidth]{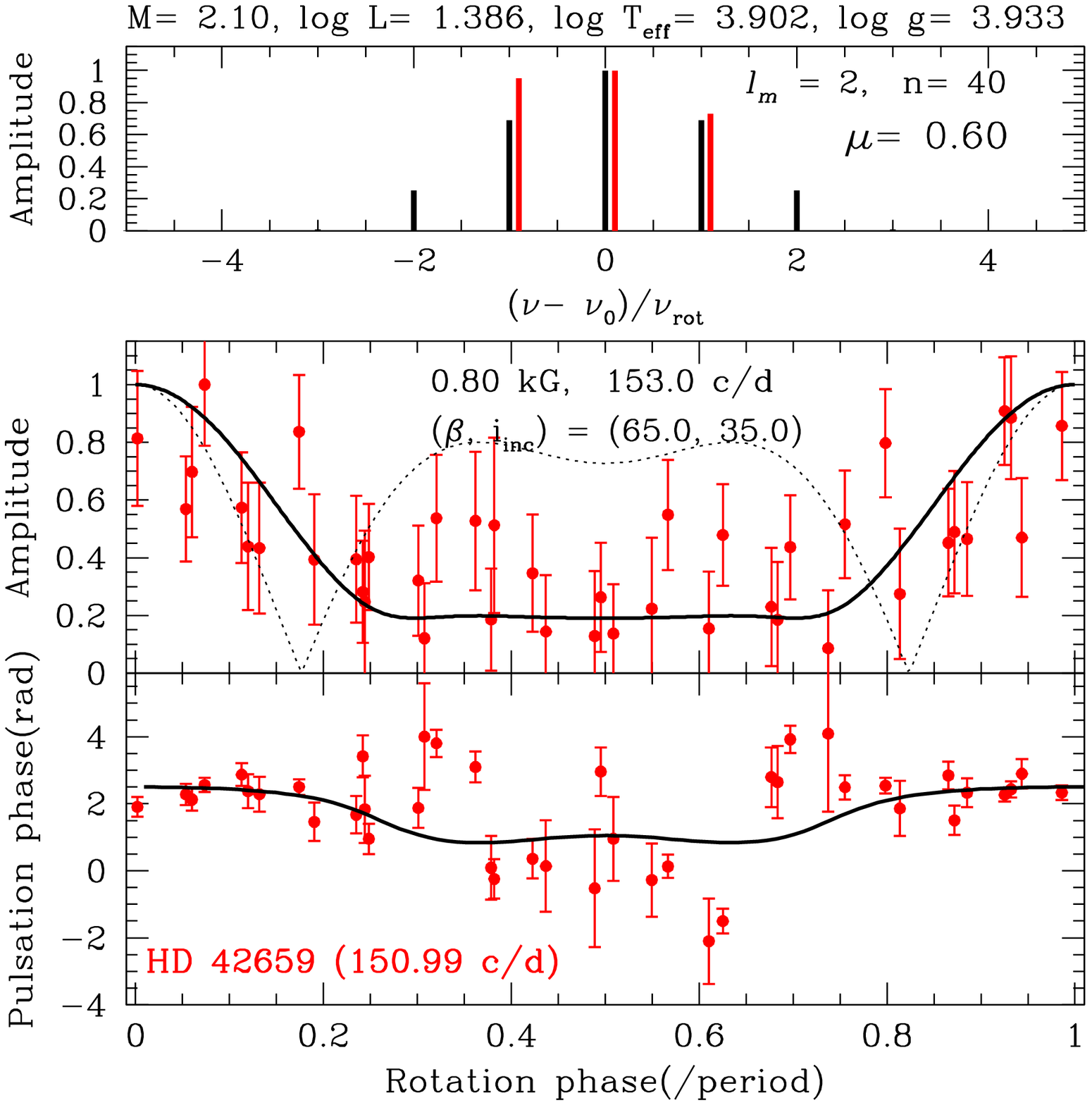}
\caption{Left: best fitting distorted dipole model to the observed amplitude and phase variations. We show a representative model, as any values of $i_{\rm inc}$, and $\beta$ which satisfy equation\,\ref{eq:OPM_trip} and Fig.\,\ref{fig:i_beta} provide good fits to the observations. Right: the best fitting distorted quadrupole model. In this case, only a small value of the polar magnetic field strength allows us to obtain a good fit as stronger fields predict smaller phase variations due to a more significant contribution from the spherically symmetric component. In both cases, the red colour represents the observations, with the black being the models. The dotted lines represent the non-distorted case in each model.}
\label{fig:model}
\end{figure*}

Our final task with the TESS data is to model the amplitude and phase variations in Fig.\,\ref{fig:phamp} following the method of \citet{saio2005}, as employed and discussed by \citet{holdsworth2016}. The magnetically distorted eigenfunction of a roAp pulsation mode is expanded using axisymmetric spherical harmonics, $Y_\ell^0$. The contribution of each $\ell$ component to the observed light variations is proportional to 
\begin{equation}
     I_\ell \equiv \int_0^{\pi/2}P_\ell(\cos\theta_L)
(1-\mu+\mu\cos\theta_L)\cos\theta_Ld(\cos\theta_L)
\end{equation}
\citep[e.g.,][]{saio2004}, where $\theta_L$ is the polar angle to the line of sight and $\mu$ is the limb darkening coefficient. A distorted dipole mode consists of odd $\ell$, in which $\ell\ge3$ components are generated from coupling with the magnetic field. However, the above equation gives $I_{\ell=3}/I_{\ell=1}=0.09$ and $I_{\ell=5}/I_{\ell=1}=-0.01$, $\dots$ (where $\mu=0.6$ is used), indicating that $\ell\ge3$ components hardly contribute to the observed light variations. Therefore, even a strongly distorted dipole mode approximately obeys the rules for a pure dipole mode. Furthermore, neither the amplitude nor the phase modulations are affected by the strength of the magnetic field. As such, we cannot constrain $i_{\rm inc}$, $\beta$ or $B_p$ (the polar magnetic field strength) by the fitting. We present in the left panel of Fig.\,\ref{fig:model} the best fitting dipole mode to the data. The choice of $B_p$ is arbitrary due to the aforementioned reasons.

We postulated, as a result of its similarity to distorted quadrupole mode pulsators \citep{holdsworth2018c}, that HD\,42659 could rather be pulsating in a distorted quadrupole mode. Therefore, we attempted to model the amplitude and phase variations accordingly. In the case of a distorted quadrupole mode, mainly the $\ell=0$ and $\ell=2$ components affect the shape of the amplitude/phase variations. This is also the case for the magnetic field. The right panel of Fig.\,\ref{fig:model} shows our best fitting model in the case of a distorted quadrupole mode.

In the case of the quadrupole mode, the polar magnetic field strength has to be sufficiently weak ($1\leq$\,kG) for the phase variations to be modelled; with a stronger field, the predicted phase variations are too small and represent a poor fit to the observations. This weak field is consistent with the mean longitudinal magnetic field strength of $0.40\pm0.05$\,kG presented in Table\,\ref{tab:params}. 

Unfortunately, the intrinsic low amplitude of the pulsation mode, coupled with the red TESS filter, mean that the observations do not have sufficient precision to allow us to conclude confidently on the degree of the mode in HD\,42659; however, the evidence we have presented leads us to favour a dipole mode for the pulsation in HD\,42659.

\section{Spectroscopy}
\label{sec:spec}

Since HD\,42659 is the only roAp star known to be in a spectroscopic binary system \citep{hartmann2015}, it provides us with the opportunity to test whether the binary companion affects the pulsations in the Ap star. To that end, we obtained 27 spectra with the High Resolution Spectrograph \citep[HRS;][]{HRS1,HRS2} mounted on the Southern African Large Telescope \citep[SALT;][]{SALT}. Observations were taken in High Resolution (HR) mode, $R\simeq65\,000$, with an integration time of 158\,s. The signal-to-noise ratio varied between about 100 and 150 in the central region of the spectrum, dependent on observing conditions. 

HRS is a dual-beam spectrograph with wavelength coverage of $3700-5500$\,\AA\ and $5500-8900$\,\AA. In the following analysis, we chose to use only the blue arm data. This removed any issues with Telluric line contamination (which is significant given the elevation of the SAAO site) and removed any systematic offsets between the blue and red arm wavelength calibrations.

The observations were automatically reduced using the SALT custom pipeline, which is based on the ESO's {\sc{midas}} pipeline \citep{pyhrs2,pyhrs3}. The data underwent standard calibrations, including flat-fielding, order extraction and wavelength calibration. Finally, the orders were merged to provide a 1-D continuous spectrum for each arm. Finally, we applied a Barycentric velocity correction to the observations.

\subsection{Radial velocity measurements}

Although the SALT pipeline provides Heliocentric radial velocities of a given spectrum, we chose to perform our own radial velocity analysis. To do this, we firstly synthesised a spectrum using {\sc{iSpec}} \citep{iSpec1,iSpec2} using the stellar parameters in Table\,\ref{tab:params}, then used this synthetic spectrum to normalise the SALT observations, also using {\sc{iSpec}}.

To derive the RV measurements, we used the cross-correlation function (CCF) technique. Each observation was cross-correlated with the synthetic spectrum, with the RV being taken when CCF is maximised. The RV, and error, were derived by a Gaussian fit to the shape of the CCF.

\begin{figure}
\includegraphics[width=\columnwidth]{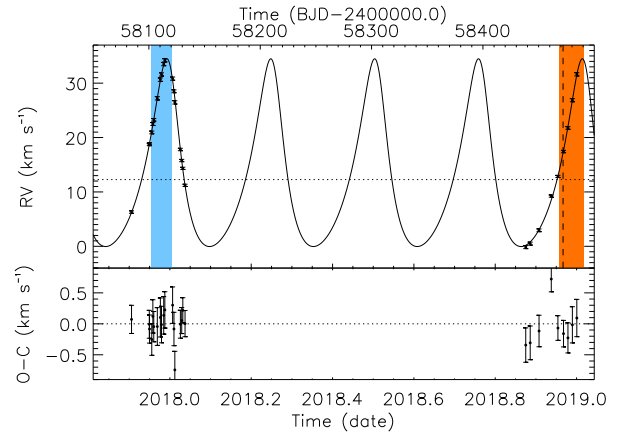}
\caption{SALT/HRS RV measurements and derived fit. The blue shading around the start of 2018 indicates the time period that SAAO photometry was obtained, with the orange shading around the start of 2019 shows the TESS data range. The vertical dashed line in the orange shading represents a time when an eclipse is expected, if the inclination allows.}
\label{fig:RV}
\end{figure}

Finally, with an RV measurement for each spectrum, we repeated the above process but removed the initial RV, then reapplied it, during the normalisation process. This served to mitigate any errors introduced to the initial RV measurement when automatically normalising the spectra. The final RV measurements are shown in Fig.\,\ref{fig:RV} and Table\,\ref{tab:RV_measures}.

\begin{table}
    \centering
    \caption{The measured RVs of HD\,42659. The time is given as BJD-2450000.0 and corresponds to the mid-point of the exposure.}
    \label{tab:RV_measures}
    \begin{tabular}{lr}
    \hline
    \multicolumn{1}{c}{Time} & \multicolumn{1}{c}{RV}\\
         & \multicolumn{1}{c}{(\kms)}\\
         \hline
8084.4168 & $6.30 \pm 0.23$ \\
8100.3845 & $18.72 \pm 0.22$ \\
8100.5691 & $18.84 \pm 0.23$ \\
8102.5908 & $20.94 \pm 0.26$ \\
8103.5884 & $22.48 \pm 0.27$ \\
8104.3742 & $23.26 \pm 0.24$ \\
8107.5808 & $27.21 \pm 0.30$ \\
8110.3382 & $30.64 \pm 0.32$ \\
8111.3365 & $31.60 \pm 0.29$ \\
8113.3312 & $33.51 \pm 0.30$ \\
8114.3221 & $34.20 \pm 0.29$ \\
8121.3213 & $30.83 \pm 0.29$ \\
8122.5325 & $28.50 \pm 0.27$ \\
8123.3153 & $26.46 \pm 0.30$ \\
8128.3025 & $17.80 \pm 0.23$ \\
8129.5067 & $15.77 \pm 0.21$ \\
8130.4944 & $14.33 \pm 0.20$ \\
8132.4855 & $11.21 \pm 0.21$ \\
8438.4564 & $-0.15 \pm 0.28$ \\
8442.4523 & $0.50 \pm 0.27$ \\
8450.4059 & $2.94 \pm 0.25$ \\
8461.3793 & $9.24 \pm 0.21$ \\
8467.3580 & $12.85 \pm 0.20$ \\
8472.3651 & $17.43 \pm 0.21$ \\
8476.3425 & $21.76 \pm 0.24$ \\
8480.3447 & $26.87 \pm 0.29$ \\
8484.3249 & $31.63 \pm 0.30$ \\
\hline
    \end{tabular}
\end{table}

We fitted the final RV measurements with the {\sc{rvlin}} code of \citet{wright2009}, and included the bootstrapping method for error analysis by \citet{wang2012}. We provided initial estimates of the system parameters based on those presented by \citet{hartmann2015} and the mass estimate from \citet{kochukhov2006}. The results of the fitting process are shown in Table\,\ref{tab:RV_fit}.

\begin{table}
\centering
\caption{Results of the {\sc{rvlin}} fit to the SALT/HRS RV data.}
\label{tab:RV_fit}
\begin{tabular}{lr}
\hline
Parameter & \multicolumn{1}{c}{Value} \\
\hline
$P$ (d) & $93.266\pm0.033$\\
$T_{\rm p}$ & $8119.52\pm0.20$\\
$K$ (km\,s$^{-1}$) & $17.26\pm0.10$\\
$e$ & $0.317\pm0.004$\\
$\omega$ ($^\circ$) & $25 \pm1$\\
$\gamma$ (km\,s$^{-1}$) & $12.27\pm0.08$\\
\hline
$f(m)$ ($10^{-3}$\,M$_\odot$) & $42.4\pm0.7$\\
$a\sin i_{\rm orb}$ (au) & $0.140\pm0.002$ \\
\hline
\end{tabular}
\end{table}

We find significantly different results for most parameters to those presented by \citet{hartmann2015}, with only the period we derive being similar. Our solution suggests a much more eccentric orbit ($e=0.317\pm0.004$ compared to $e=0.146\pm0.027$), and a much larger velocity semi-amplitude. When inspecting figure 3 of \citet{hartmann2015}, it is noted that their orbital phase coverage is not complete at the very bottom of the velocity curve which perhaps has led to these discrepancies. The difference in these two parameters has an effect on the minimum mass of the companion, which in our case is larger at $0.69\pm0.01$\,M$_\odot$. We see no indication of the secondary star in the spectra which, when considering the mass function and magnitude ratios, implies that the companion is between mid-F and early-K. In light of this, we double checked the light curve to ensure we had not pre-whitened or neglected any other signal which could be attributed to the companion. None was found. 

With the binary solution, we were able to predict the epochs of eclipse events using the {\sc{exofast}} routine of \citet{eastham2013}. The TESS observations coincide with one possible epoch, as denoted by the vertical dashed line in Fig.\,\ref{fig:RV}. However, we see no evidence of an eclipse in the light curve.

\section{Discussion and conclusions}

The ground based photometric observations of HD\,42659 do not, unfortunately, provide significantly more information than was previously known. However, the TESS and spectroscopic data do add to our understanding of the roAp star, and the system as a whole.

The TESS photometric observations are the first data which allow us to measure the rotation period of this star. First done by \citet{balona2019}, we have refined the fit here to arrive at a rotation period of $2.66265\pm0.00025$\,d. This finding allows us to shed some light on the number of nights which \citet{martinez1994} failed to measure a pulsation; their data, as shown in their table 5 and figure, are separated such that pulsation maximum was often not observed, thus the pulsation was below their detection limit. It was fortunate that they persevered and obtained good rotation phase coverage of the star, although they did not know what that was at the time.

The TESS data have also allowed us to provide the first models of the pulsation in HD\,42659. We have presented two model fits to the data and have a preference for the dipole fit as the quality of the fit is better (with a reduced $\chi^2$ of 4.2 rather than 5.8 for the quadrupole model). However, it is only when considering a distorted quadrupole mode that we can provide estimates for the geometry and polar magnetic field strength in the star.

 Using the stellar rotation period derived here, we can estimate an equatorial rotation velocity of $v_{\rm eq}=55\pm6$\,\kms\ using the value of the radius, calculated with the standard equation, in Table\,\ref{tab:params}. This result, and knowledge of the $v\sin i_{\rm inc}$ from \citet{elkin2008}, allows us to independently estimate $i_{\rm inc}$ at $20\pm3^\circ$. This value is in agreement with the pulsation modelling results presented earlier, and is also in agreement with the results of equation\,(\ref{eq:OPM_trip}) when considering a dipole mode. We highlight, on Fig.\,\ref{fig:i_beta}, the estimated values of $i_{\rm inc}$ from this discussion.
 
 We do not assume that the orbital axis is aligned with rotation axis of the Ap star. If our estimate of $i_{\rm inc}=20\pm3^\circ$ is correct, and if the orbital and rotation axes are aligned (i.e. $i_{\rm inc}=i_{\rm orb}$), this leads to a secondary mass range of $M_2\simeq2.4-4.0$\,M$_\odot$. Since we see no sign of a secondary component in the spectra (see end of Sec.\,\ref{sec:spec}), an orbital inclination angle of $\gtrapprox35^\circ$ is required, giving a secondary mass range of $M_2\simeq0.7-1.4$\,M$_\odot$.
 
 As discussed earlier, this star has previously been identified as a possible distorted quadrupole pulsator from its position in the $T_{\rm eff}-\nu L/M$ plane \citep[see][for details]{saio2014} and its proximity to other roAp stars pulsating with distorted quadrupole modes. All of these stars pulsate above their theoretical acoustic cutoff frequency, $\nu_{\rm ac}$. This theoretical limit can be calculated by scaling from the solar value such that
 \begin{equation}
     \frac{\nu_{ac}}{\nu_{ac,\odot}} = \frac{M/{\rm M_\odot}(T_{\rm eff}/{\rm T}_{\rm eff,\odot})^{3.5}}{L/{\rm L}_\odot},
 \end{equation}
 where $\nu_{{\rm ac},\odot}$ is taken to be 458\,\cd\ \citep{jimenez2011}. Using values from Table\,\ref{tab:params}, we find $\nu_{\rm ac}$ is about $100\pm10$\,\cd. This is significantly lower than the measured pulsation mode frequency ($\sim151$\,\cd), thus confirming that HD\,42659 is a roAp star pulsating at a super-critical frequency. Since linear, non-adiabatic models show that the $\kappa$-mechanism can only excite modes close to, but not beyond, the acoustic cutoff frequency, it is possible that excitation mechanism is not driving the pulsation in HD\,42659. However, the models do not directly account for the magnetic field which plays a significant role in these super-critical pulsators. Alternatively, \citet{cunha2013} showed that models where convection is not suppressed by the magnetic field some super-critical pulsations could be driven. If this is the case here, turbulent pressure could be the driving mechanism for the pulsation in HD\,42659.
 
HD\,42659 is currently the only roAp known to be in a relatively short, spectroscopic, orbit. Other roAp stars, such as $\alpha$\,Cir and $\gamma$\,Equ \citep{scholler2012}, do have companions, but these are in wide orbits. HD\,42659 allows us to directly probe the influence of binarity on the roAp phenomenon, as suggested by \citet{hartmann2015}. The TESS data, and our simultaneous spectra, allow us to do this for the first time for a roAp star. However, there are Ap stars in close binaries that have been studied in this way. Recently, \citet{skarka2019} studied HD\,99458 both spectroscopically and photometrically. They found a primary Ap star showing $\delta$\,Sct pulsations (i.e., low-overtone p-modes) in an eclipsing binary with an M-dwarf companion. Clearly the $\delta$\,Sct pulsations were not suppressed, but the authors did not investigate this to its full potential. Furthermore, TESS observations are uncovering more Ap stars in short-period eclipsing binary systems, of which there is currently a dearth \citep[e.g.][and references within]{mathys2017}, providing more opportunities to investigate these systems.

\begin{figure}
\includegraphics[width=\columnwidth]{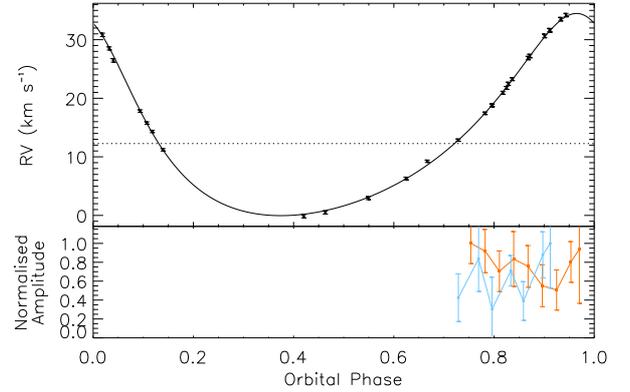}
\caption{Top: phased fit of the binary model to the RV measurements. Bottom: normalised (to the respective maximum value) photometric amplitudes of the pulsation mode averaged over the rotation period. Blue circles represent the SAAO results, with orange squares being the TESS results. The zero point for the phases is chosen to be the time of periastron passage, $T_{\rm p}$.}
\label{fig:RV_photo}
\end{figure}

The TESS data cover just under a quarter of the binary orbit, and occur on the approach to periastron passage. The orange shading in Fig.\,\ref{fig:RV} shows this visually. For a clearer view of this, we present a phase folded RV curve in Fig.\,\ref{fig:RV_photo} and below this we plot the normalised average amplitude of the pulsation, as calculated over a rotation cycle. We plot both the SAAO and TESS results in this way. These amplitudes cannot be directly compared, and are normalised to their respective maximum.

The SAAO amplitude results are inconclusive due the the pulsation S/N and the gaps in the data, which mean we are sampling different pulsation amplitudes due to the rotation. The TESS data do show a trend of $-1.0\pm0.7\,\muup$mag\,\cd\ which is not statistically significant. Therefore, it is not possible to conclude whether the binary companion is affecting the pulsation mode amplitude.

It is unfortunate that the TESS data cover such a small fraction of the orbital phase curve, but this adds further motivation to obtain high-cadence, high-quality photometric observations of HD\,42659 at other orbital phases, and especially at apastron where we currently have no photometric data. Due to the low amplitude of the mode, and the need for high-precision data, this target is a good candidate for the Whole Earth Telescope \citep{sullivan2001,provencal2014}.

TESS observations are enabling us to find new roAp stars, and reclassify known Ap stars which were thought to be constant from ground-based observations as variable \citep{cunha2019,balona2019}. Those targets are prime candidates to perform long-term spectroscopic studies to search for binary companions which may affect any pulsation mode amplitudes.

\section*{Acknowledgements}

We thank the anonymous referee for a useful comments and suggestions. DLH and DWK acknowledge financial support from the STFC via grant ST/M000877/1. 
This paper uses observations made at the South African Astronomical Observatory (SAAO). 
Some of the observations reported in this paper were obtained with the Southern African Large Telescope (SALT) under programmes 2018-2-SCI-016 and 2017-2-SCI-013, PI: Holdsworth. This paper includes data collected by the TESS mission. Funding for the TESS mission is provided by the NASA Explorer Program. Funding for the TESS Asteroseismic Science Operations Centre is provided by the Danish National Research Foundation (Grant agreement no.: DNRF106), ESA PRODEX (PEA 4000119301) and Stellar Astrophysics Centre (SAC) at Aarhus University.

\bibliography{Holdsworth-refs}
\label{lastpage}
\end{document}